\documentclass[11pt,twoside]{article}

\usepackage{asp2006}
\usepackage{graphicx}

\begin{document}
\title{The Evolution of Galaxy Dust Properties for $1 < z < 2.5$}   

\author{Stefan Noll, Daniele Pierini, Maurilio Pannella, and Sandra Savaglio}

\affil{Max-Planck-Institut f\"ur extraterrestrische Physik, Giessenbachstr., 
85748 Garching, Germany} 

\begin{abstract} 
Fundamental properties of the extinction curve, like the slope in the 
rest-frame UV and the presence/absence of a broad absorption excess centred 
at 2175\,\AA{} (the UV bump), are investigated for a sample of 108 massive, 
star-forming galaxies at $1 < z < 2.5$, selected from the FDF Spectroscopic 
Survey, the K20 survey, and the GDDS. These characteristics are constrained 
from a parametric description of the UV spectral energy distribution of a 
galaxy. It turns out that the sample galaxies host dust producing extinction 
curves with properties in between those of the Small and Large Magellanic 
Clouds (SMC and LMC, respectively). LMC-like extinction curves, which exhibit 
a UV bump, are mainly found among highly-reddened, UV-ultraluminous galaxies 
at $z \sim 2.4$ and highly-reddened, near-IR-bright, star-forming galaxies at 
$z \sim 1.2$. We discuss star-formation rates, total stellar masses, the 
morphology, and the chemical properties of our sample galaxies with respect 
to possible explanations for the different extinction curves. 
\end{abstract}

\section{Introduction}

Nearby starburst galaxies probably contain dust with an extinction curve 
lacking a broad bump centred at 2175\,\AA{} \citep{noll:CAL94}. This is 
typical of the Small Magellanic Cloud (SMC), whose harsh environments 
(i.e.~strong radiation fields and shocks) and/or low metallicity do not allow 
a large presence of those grains that are proposed as the carriers of the UV 
bump \citep{noll:GOR03}. However, the supershell region surrounding 30\,Dor 
in the Large Magellanic Cloud (LMC) contains such carriers, which are 
ubiquitous in the diffuse interstellar medium (ISM) of the Milky Way.  

Are there star-forming galaxies at high redshift exhibiting the dust 
absorption feature at 2175\,\AA{} in their spectra? While for most objects 
the answer to this question is no or at least doubtful 
\citep[e.g.,][]{noll:VIJ03, noll:WAN04, noll:YOR06}, \citet{noll:NOL05} find 
that the majority of reddened UV-luminous galaxies at $2 < z < 2.5$ show 
clear evidence of the 2175\,\AA{} bump. These galaxies seem to experience 
strong dust reddening \citep{noll:NOL04} and seem to be massive and metal 
rich \citep{noll:MEH02, noll:DAD04, noll:SHA04}. In order to investigate 
whether the carriers of the 2175\,\AA{} feature exist in other galaxy 
populations at different cosmic times, we extend our analysis of rest-frame 
UV spectra to additional UV-luminous and massive galaxies at $1 < z < 2.5$ 
(Noll et al., in prep.).

\section{The Spectroscopic Sample}

In addition to the 34 galaxies at $2 < z < 2.5$ with $R_{\rm Vega} < 24.8$ 
already studied by \citet{noll:NOL05}, we select 32 $1 < z < 2$ objects with 
$R < 24$ from the $I$-limited FORS Deep Field (FDF) Spectroscopic Survey 
\citep{noll:NOL04}. This sample is complemented by 34 star-forming galaxies 
at $1 < z < 2.3$ selected from the spectroscopic catalogue of the K20 Survey 
in the Chandra Deep Field South and a field around the quasar 0055-2659 
\citep{noll:CIM02, noll:MIG05}. Besides $K_{\rm s} < 20$ all objects but two 
have $R < 24$. Finally, we consider eight star-forming galaxies at 
$1.5 < z < 2.3$ from the $I$ and $K$-limited Gemini Deep Deep Survey 
\citep[GDDS,][]{noll:ABR04}, all with $R < 24.7$ and $K_{\rm s} \la 21.0$.

\section{Analysis}

In order to study the UV spectral energy distributions of our sample galaxies 
we use parameters based on power-law fits of the form 
$f(\lambda) \propto \lambda^{\gamma}$ to sub-regions of the UV continuum. The 
UV-bump indicator $\gamma_{34} = \gamma_3 - \gamma_4$ \citep{noll:NOL05} is 
based on continuum slope measurements in the wavelength ranges 
$1900 - 2175$\,\AA{} ($\gamma_3$) and $2175 - 2500$\,\AA{} ($\gamma_4$), 
respectively. The ``classical'' UV reddening measure $\beta$ is taken for 
$1250 - 1750$\,\AA{} \citep{noll:CAL94, noll:LEI02}. Due to a different 
accessible wavelength range we take $\beta_{\rm b}$ as a reddening measure 
for galaxies at $z < 2$. This parameter is derived at $1750 - 2600$\,\AA{} 
excluding $1950 - 2400$\,\AA{}, i.e.~the range of the 2175\,\AA{} feature. 
Wavelength regions affected by strong spectral lines are also excluded from 
the fitting procedure for all parameters 
\citep[see][]{noll:CAL94, noll:NOL05}.

\section{Results}

\begin{figure}[!ht]
\centering 
\includegraphics[width=8cm,clip=true]{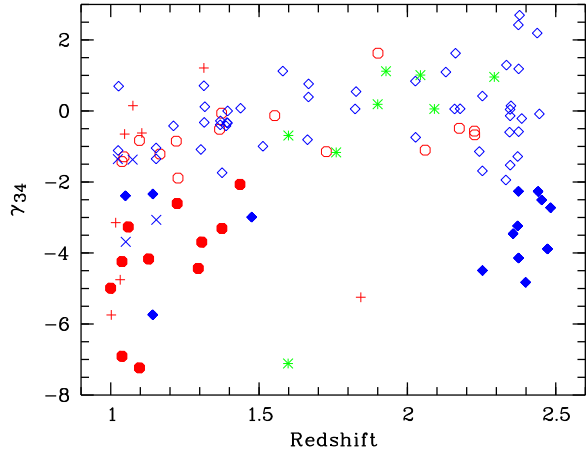}
\caption[]{The UV-bump indicator $\gamma_{34}$ versus redshift for 108 
actively star-forming galaxies at $1 < z < 2.5$ selected from the FDF 
Spectroscopic Survey (lozenges and $\times$), the K20 Survey (circles and 
$+$), and the GDDS (asterisks). Galaxies without a UV-continuum slope 
derivation ($\times$, $+$, and asterisks) are no more considered in 
Figs.~\ref{noll:fig2} and \ref{noll:fig3}.}
\label{noll:fig1}
\end{figure}

\begin{figure}[!ht]
\centering 
\includegraphics[width=8cm,clip=true]{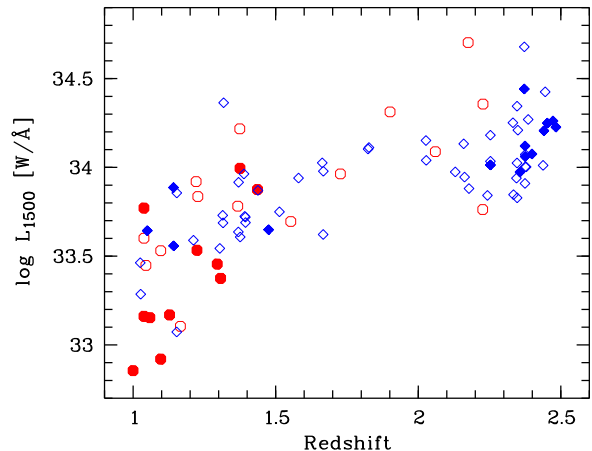}
\caption[]{Luminosity at 1500\,\AA{} versus redshift for 88 galaxies of the 
FDF Spectroscopic Survey (lozenges) and the K20 survey (circles). Galaxies 
with $\gamma_{34} < -2$ (i.e.~with a prominent 2175\,\AA{} feature) are 
marked by filled symbols.}
\label{noll:fig2}
\end{figure}

Fig.~\ref{noll:fig1} shows the UV-bump indicator $\gamma_{34}$ versus 
redshift for our spectroscopic sample of 108 actively star-forming galaxies 
at $1 < z < 2.5$. Galaxies with strong 2175\,\AA{} features, 
i.e.~$\gamma_{34} < -2$, are mainly found at $2.3 < z < 2.5$ (38\% of the FDF 
galaxies in this redshift range) and $1 < z < 1.5$. Considering that about 
29\% of the objects in the whole photometric FDF sample at $1 < z < 1.5$ with 
$R < 24$ show $K_{\rm s} < 20$ \citep{noll:HEI03, noll:GAB04}, we can combine 
the FDF and K20 samples and estimate that about 25\% of the $1 < z < 1.5$ 
galaxies with $R < 24$ have $\gamma_{34} < -2$. 87\% of these galaxies exhibit 
$K_{\rm s} < 20$. Among the $1 < z < 1.5$ galaxies, only objects with 
$z < 1.2$ show very strong 2175\,\AA{} features as indicated by 
$\gamma_{34} < -5$. At $1.5 < z < 2.2$ only two out of 26 galaxies indicate 
$\gamma_{34} < -2$. In one case the redshift is uncertain and in the other 
case $R = 24.4$, which does not fulfil the $R < 24$ criterion. The latter 
object suggests that $1.5 < z < 2.2$ galaxies with relatively strong UV bumps 
could pop up just at fainter luminosities than covered by the selection 
limit $R = 24$ (see Fig.~\ref{noll:fig2}). Most sample galaxies with 
$\gamma_{34} < -2$ at $z \sim 1.1$ would not have been detected, if they had 
been at $1.5 < z < 2.2$. On the other hand, there seems to be a real lack of 
high-luminosity objects with $\gamma_{34} < -2$ at these redshifts, since
objects like those found at $z \sim 2.4$ should be easily detectable at 
$1.5 < z < 2.2$ (see Fig.~\ref{noll:fig2}).    

\begin{figure}[!ht]
\centering 
\includegraphics[width=6cm,clip=true]{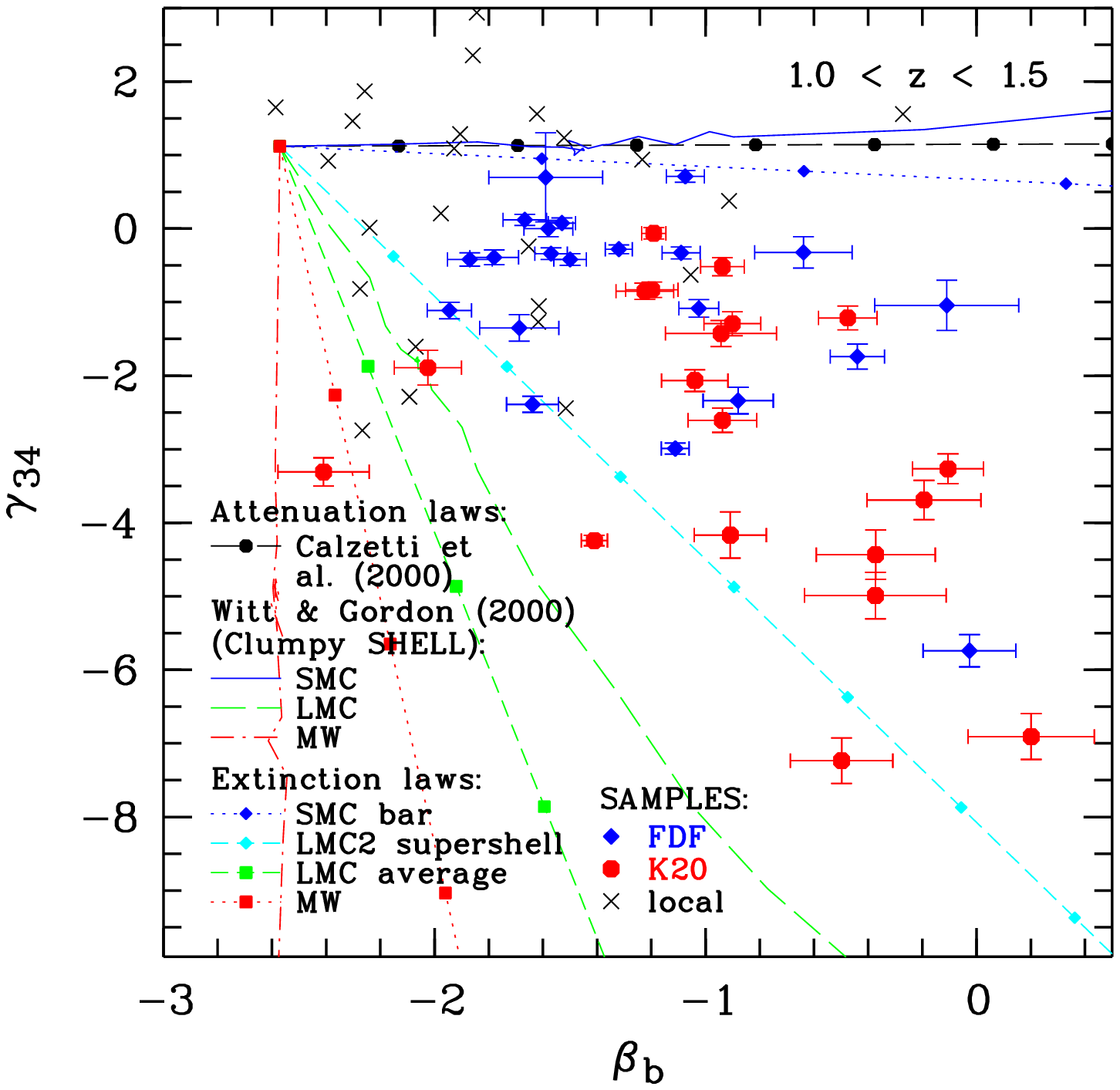}
\includegraphics[width=6cm,clip=true]{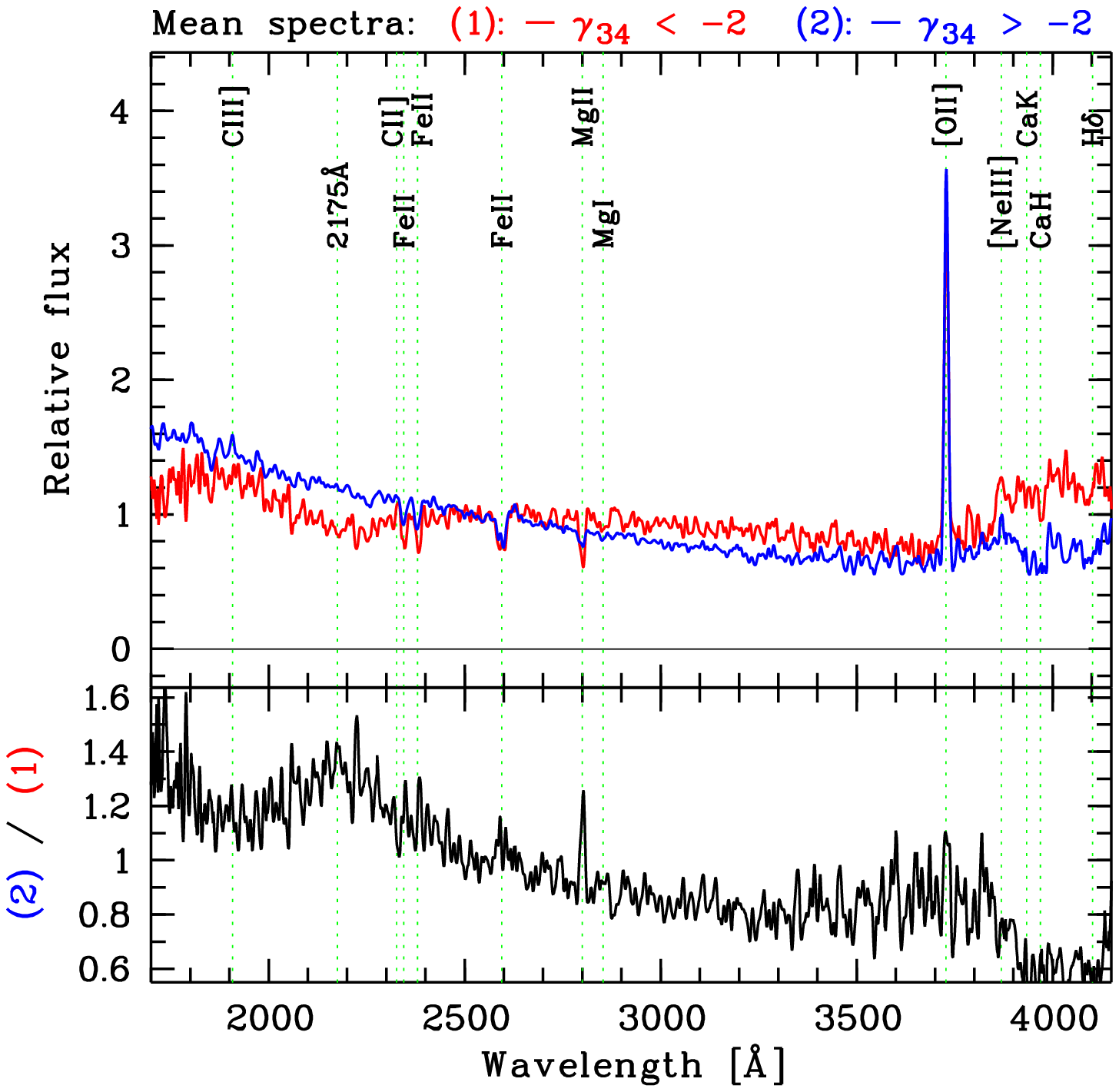}
\caption[]{{\em Left:} The UV-bump indicator $\gamma_{34}$ versus the 
reddening measure $\beta_b$ for our $1 < z < 2.5$ FDF (lozenges), and K20 
galaxies (circles), and a comparison sample of 24 local starburst galaxies 
(crosses) observed with IUE \citep[see][]{noll:NOL05}. The diagram also shows 
different dust attenuation models (see legend) for a Maraston (2005) stellar 
population synthesis model with Salpeter IMF, continuous SF, an age of 
100\,Myr, and solar metallicity. The symbols are plotted in intervals of 
$\Delta E_{B-V} = 0.1$. {\em Right:} Comparison of $K_{\rm s}$-weighted 
composite spectra of our $1 < z < 1.5$ galaxies exhibiting $\gamma_{34} > -2$ 
(blue or dark grey) and $\gamma_{34} < -2$ (red or grey), respectively 
({\em top}). The ratio of both composites, normalised at 2400 -- 2570\,\AA{}, 
is also shown ({\em bottom}).}
\label{noll:fig3}
\end{figure}

\citet{noll:NOL05} find that the UV continua of UV-luminous galaxies at
$2 < z < 2.5$ suggest effective extinction curves ranging from those typical
of the SMC to those typical of the LMC. Fig.~\ref{noll:fig3} shows that this 
also seems to be true for galaxies at $1 < z < 1.5$. The average 
$\gamma_{34}$ decreases with increasing UV-reddening parameter $\beta_{\rm b}$ 
as expected for extinction curves showing a significant UV bump. For 
$\beta_{\rm b} < -1.5$ the $K_{\rm s}$-weighted average 
$\langle \gamma_{34} \rangle = -0.44 \pm 0.18$, while for 
$\beta_{\rm b} > -0.5$ $\langle \gamma_{34} \rangle = -3.48 \pm 0.51$. 

Fig.~\ref{noll:fig3} also shows a comparison of the $K_{\rm s}$-weighted 
composite spectra of $1 < z < 1.5$ galaxies with $R < 24$ having a weak 
($\gamma_{34} > -2$) and a strong ($\gamma_{34} < -2$) 2175\,\AA{} absorption
feature, respectively. The presence of the feature in the more reddened 
composite with $\gamma_{34} < -2$ is evident. Moreover, as indicated by the 
stronger Balmer/4000\,\AA{} break \citep[e.g.,][]{noll:BAL99} this spectrum 
seems to have an important contribution of an intermediate-age stellar 
population (i.e.~$10^2 - 10^3$\,Myr) and/or to show an about three times 
older stellar population than the composite of galaxies with weak or absent 
UV bump. Finally, Fig.~\ref{noll:fig3} exhibits a stronger Mg\,II absorption 
in relation to the strength of the Fe\,II absorption for the composite of the 
galaxies with $\gamma_{34} < -2$. This finding is also consistent with the 
above age estimates, whether the nature of the Mg\,II absorption is 
interstellar only (chemical enrichment of Mg in the ISM) or, also, partly 
stellar (composition of the stellar population).   
 
\begin{figure}[!ht]
\centering 
\includegraphics[width=6cm,clip=true]{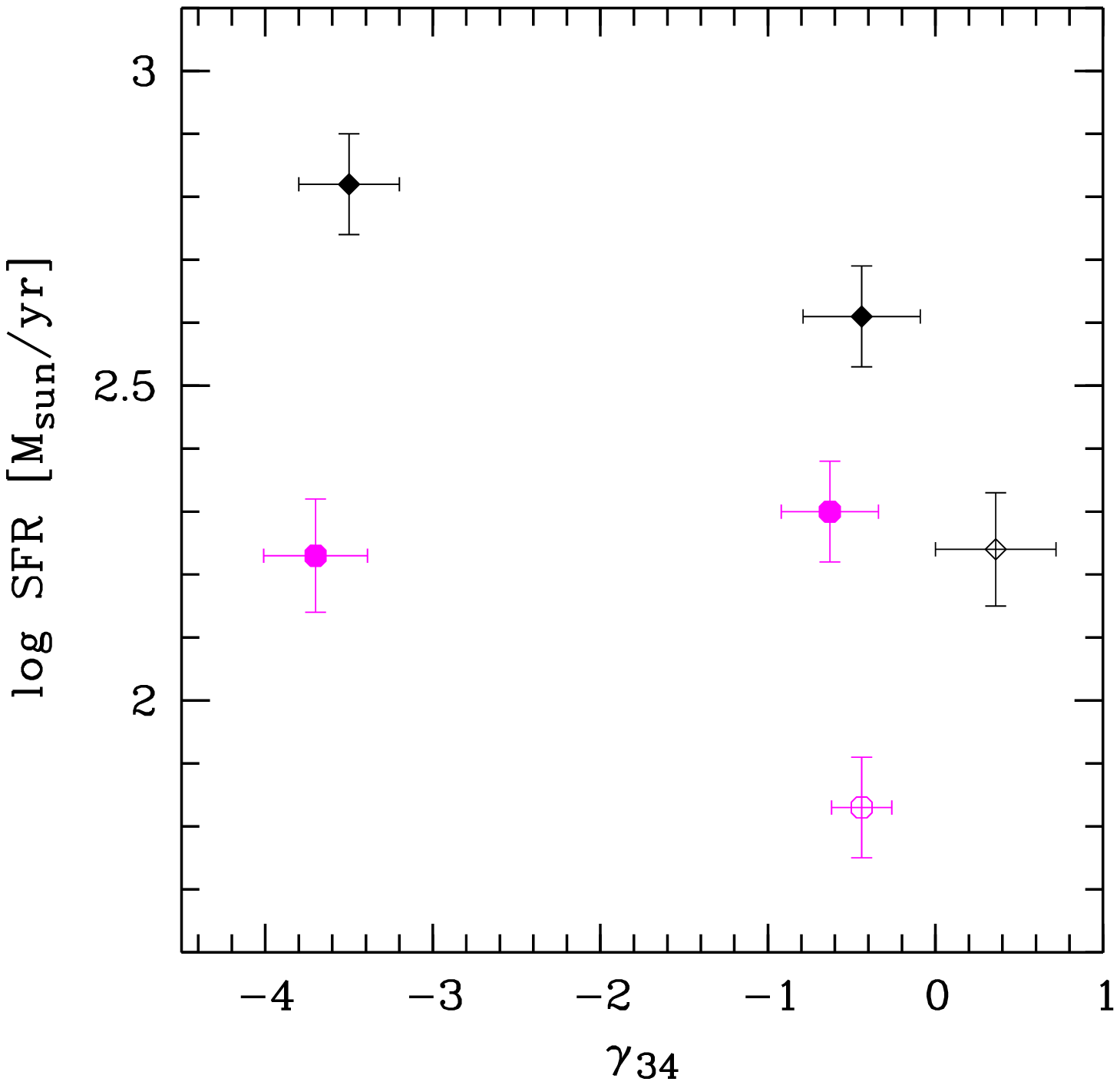}
\includegraphics[width=6cm,clip=true]{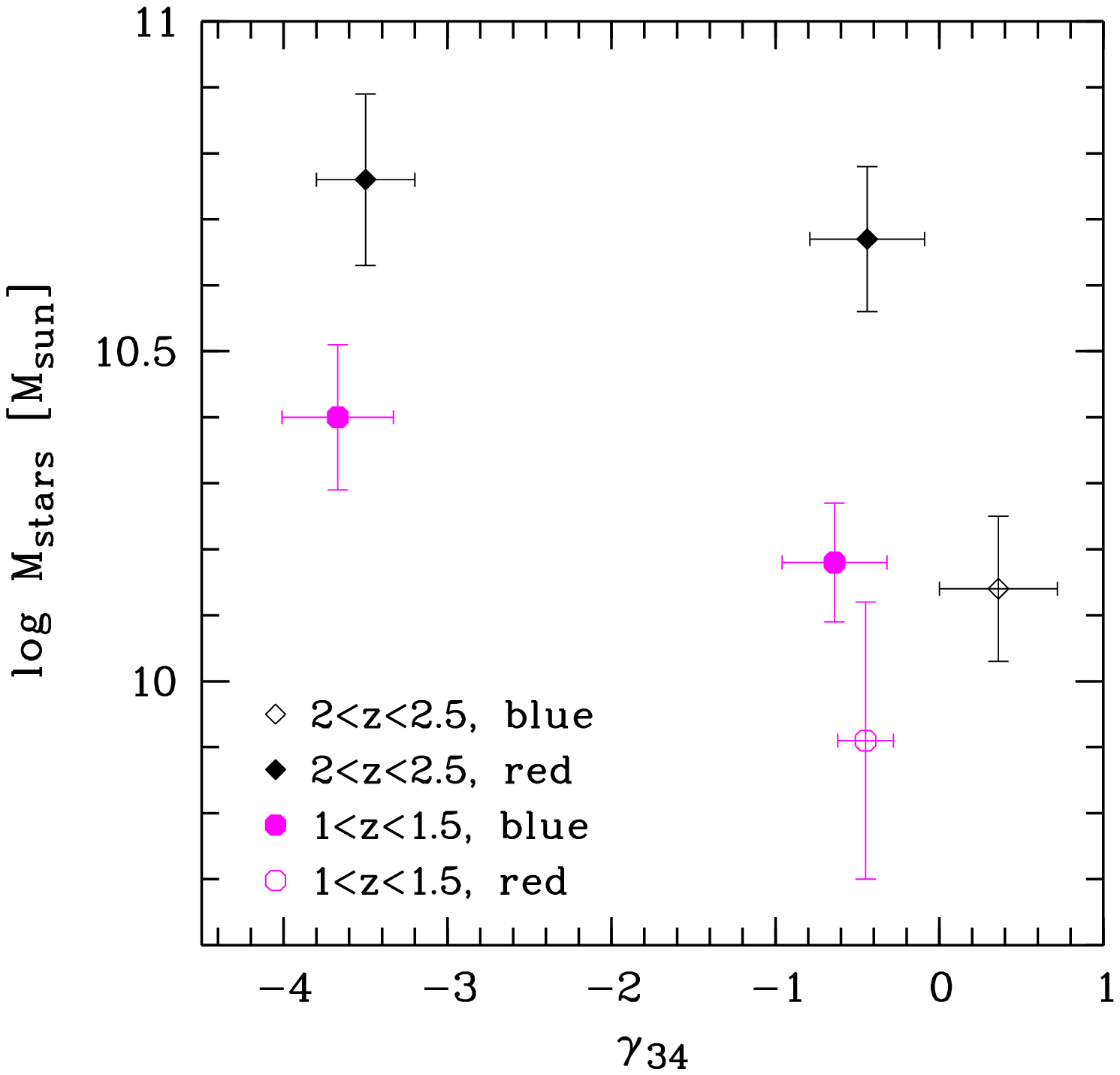}
\caption[]{SFR ({\em left}) and total stellar mass ({\em right}) versus the 
UV-bump proxy $\gamma_{34}$ for six subsamples. Lozenges and circles indicate 
the FDF $2 < z < 2.5$ and the FDF+K20 $1 < z < 1.5$ galaxies, respectively. 
Open symbols mark the subsamples with low reddening ($\beta < -0.4$; 
$\beta_{\rm b} < -1.5$), while the highly reddened galaxies are represented 
by filled symbols. The errors indicated are mean errors.}
\label{noll:fig4}
\end{figure}

From fits (the UV-bump range excluded) of the spectra of our sample galaxies 
with stellar population synthesis models of \citet{noll:MAR05} reddened using 
the ``Calzetti law'' \citep{noll:CAL00} we obtain average star-formation
rates (SFRs) for different subsamples as indicated in Fig.~\ref{noll:fig4}.   
For both redshift ranges the ``red'' subsamples (i.e.~$\beta < -0.4$ for 
$2 < z < 2.5$ and $\beta_{\rm b} < -1.5$ for $1 < z < 1.5$) have higher 
average SFRs than the ``blue'' subsamples due to larger dust reddening. 
Moreover, the $1 < z < 1.5$ galaxies exhibit lower SFRs (by a factor of 2 to 
4) than the $2 < z < 2.5$ galaxies. For the total stellar masses 
\citep{noll:DRO05, noll:PAN06} Fig.~\ref{noll:fig4} shows similar trends as
those for the SFR. Bluer and lower-redshift galaxies tend to have lower 
masses. The latter trend is in agreement with the ``downsizing'' scenario
\citep{noll:COW96}. Estimates of the metallicity using different spectral 
features for $2 < z < 2.5$ \citep{noll:MEH02, noll:MEH05} and the 
mass-metallicity relation of \citet{noll:SAV05} for $1 < z < 1.5$ (because of 
the lack of suitable spectral features in the accessible wavelength range) 
suggest typical values close to solar for at least the red subsamples in both 
redshift ranges. Hence, for constant UV-continuum reddening SFR, mass, and 
metallicity do not obviously correlate with the strength of the UV bump. 

We analyse the morphology of our sample galaxies from red Hubble ACS images
using GIM2D \citep{noll:SIM99} for a PSF-convolved S\'ersic-profile fitting 
and the CAS parametrisation of \citet{noll:CON00, noll:CON03}. For 
$1 < z < 1.5$ the red galaxies tend to have larger radii, and lower S\'ersic 
and concentration indices than the blue galaxies. The latter two parameters 
also seem to be lower for red galaxies with stronger UV bumps. For the 
S\'ersic index, e.g., we find $0.54 \pm 0.25$ ($\beta_{\rm b} > -1.5$ and 
$\gamma_{34} < -2$) and $0.98 \pm 0.23$ ($\beta_{\rm b} > -1.5$ and 
$\gamma_{34} > -2$), respectively. The parameters measured and a visual 
inspection of the images imply that particularly $1 < z < 1.5$ galaxies 
with relatively strong UV bumps are often disc galaxies. At $2 < z < 2.5$ 
there are obviously similar trends in the morphological parameters (at least 
for the relation between blue and red galaxies), but less significant. A 
difference is the higher ``clumpiness'' of the light distributions of the red 
galaxy sample. From a visual inspection of these galaxies we learn that at 
least half of the objects exhibit multiple components due to a patchy 
distribution of young stars and gas and/or (major) merger events.

\section{Conclusions}

We have constrained properties of the extinction curves for 108 UV-luminous 
galaxies at $1 < z < 2.5$. The strongest evidence for the dust absorption 
feature at 2175\,\AA{} (which can be as strong as in the LMC) comes from 
objects around $z \sim 1.1$ and $z \sim 2.4$.

At $2 < z < 2.5$ galaxies with relatively strong UV bumps are highly 
dust-enshrouded, i.e.~there is a high neutral clouds' covering fraction in 
the direction towards the observer \citep[as suggested by the study of strong 
interstellar UV absorption lines;][]{noll:NOL05}. This implies that the 
carriers of the 2175\,\AA{} feature are protected from the strong and hard 
radiation fields and/or shocks in these ultraluminous galaxies by dust 
self-shielding due to high dust column densities. At $1 < z < 1.5$ the 
galaxies with relatively strong UV bumps are less massive, show lower SFRs, 
have similar sizes, contain a relatively high fraction of intermediate-age 
stars (including carbonaceous-dust producing AGB stars), and tend to be disc 
galaxies. Lower SFR densities could mean that the carriers of the 2175\,\AA{} 
feature are exposed to less harsh conditions in the ISM, which can ease the 
need for self-shielding.

\acknowledgements This research was supported by the German Science 
Foundation (DFG, SFB 375).

\end{document}